\documentstyle[svcon]{report}
\begin{document}
\pagenumbering{arabic}
\chapter{Decaying Neutrinos and the Flattening of the Galactic Halo}
\chapterauthors{D.W. Sciama}
\noindent
S.I.S.S.A., I.C.T.P., Strada Costiera 11, 34014 Trieste, Italy\\
Nuclear and Astrophysics Laboratory, University of Oxford\\
e-mail: sciama@sissa.it

\begin{abstract}
The recently constructed Dehnen-Binney set of mass models for the
Galaxy is used to show that the decaying neutrino theory for the ionisation
of the interstellar medium (Sciama 1990a, 1993) requires the neutrino
halo of the Galaxy to be as flattened as is observationally permitted
(axial ratio
${q=0.2}$ or shape E8). The argument involves an evaluation
of the contribution of red-shifted decay photons from the cosmological
distribution of neutrinos to the extragalactic diffuse background at ${1500
\AA}$.
This contribution must be as large as is observationally permitted. These two
requirements depend on the decay lifetime ${\tau}$ in potentially
conflicting ways.
For consistency to be achieved $\tau$ must lie within 30 per cent of
$10^{23}$ seconds.
\end{abstract}

\section{Introduction}
It gives me great pleasure to dedicate this paper to Engelbert Schucking, whose
subtle mind has illuminated many problems in cosmology and general
relativity. I hope that he enjoys the way in which, because of the
pervasiveness of neutrinos,
cosmology and galactic astronomy become interdependent in the decaying
neutrino theory
for the ionisation of the interstellar medium (Sciama 1990a, 1993). I shall
give an example of
this interdependence here. My discussion also exemplifies two other
useful features of the decaying neutrino theory. Firstly the theory leads
to specific and testable predictions concerning the configuration of various
matter distributions. For example, it predicts that, if the ionisation of
hydrogen in an opaque region of the Galaxy is mainly due to decay
photons, then the resulting electron density will be independent of
the neutral hydrogen density in the region. There is observational
evidence in support of this prediction (Sciama 1990b, 1997).
The second feature is that various observations have reduced the
domain of validity of the theory to a small region of its parameter space.
For example, the energy $E_{\gamma}$ of a decay photon in the rest frame
of the decaying neutrino is constrained with a precision of one per cent
($E_{\gamma}=13.7\pm0.1\; eV$). While this feature may eventually lead to
the demise of the
theory, it has so far managed to survive.

In this paper I provide another example of these two features, which is
based on a new set of comprehensive mass models of the Galaxy (Dehnen and
Binney 1997).
I will demonstrate that, if the decaying neutrino theory is correct, the
neutrino halo of our
Galaxy must be flattened to the maximum extent allowed by these models, that
is, with an axial ratio $q=0.2$, corresponding to a shape factor E8
(where q is related to En by $q=1-n/10)$. Associated with this result is a
constraint imposed on the neutrino decay lifetime $\tau$, whose value is
required to be close to
$10^{23}$ seconds.
The argument leading to these conclusions is based on the following
observational
results in addition to those underlying the Dehnen-Binney models:

\vskip 8.25pt
{\it (i)} Pulsar dispersion data imply that the free election density in
the intercloud medium within one
kiloparsec of the sun lies in the range $0.04-0.06 \;{\rm cm}^{-3}$
(Reynolds 1990,
Sciama 1990b) and these free electrons have a scale height ${\sim} 1$ kpc
(Reynolds 1991, Nordgren et al 1992, Taylor and Cordes 1993).

\vskip 8.25pt
{\it (ii)} H${\alpha}$ data imply that there are $4{\rm x}10^6$ hydrogen
ionisations per ${\rm cm}^2$ per sec along a column at the
sun perpendicular to the galactic plane (Reynolds 1984).

\vskip 8.25pt
{\it (iii)} The isotropic extragalatic photon flux at $1500 {\AA}{\sim} 300
\pm80$  photons cm$^{-2}$  sec$^{-1}$ ster$^{-1}\;{\AA}^{-1}$
(continuum units or CU) (Henry and Murthy 1993, Witt and Petersohn 1994).
This value is still somewhat controversial (compare Bowyer 1991 with Henry
1991). In fact a significantly smaller value would rule out the decaying
neutrino theory, as we shall see.

\section{The Neutrino Density near the Sun}
The neutrino density $n_{\nu}(0)$ near the sun, which we are assuming to be
mainly responsible for the free electron density $n_e$ in opaque regions of
the intercloud medium, will be given in ionisation equilibrium by
$$\frac {n_{\nu}(0)}{\tau}=\alpha n_e^2\;,$$
where $\alpha$ is the hydrogen recombination coefficient excluding
transitions directly to the ground state.
There is, however, a danger that the decaying neutrino theory may lead to a
value for $n_{\nu}(0)$ which is larger than is permitted by the
Dehnen-Binney mass models. We therefore immediately adopt the smallest
observationally allowed values for $\alpha$ and $n_e$, namely
$2.6$x$10^{-13}$cm$^3$sec$^{-1}$ (corresponding to the reasonable electron
temperature of $10^4$ K (Osterbrock 1989)) and $0.04$ cm $^{-3}$
respectively. Then
$$n_{\nu}(0)=4.16{\rm x}10^7 {\tau}_{23}\;{\rm cm}^{-3}\;,$$
where ${\tau}=10^{23}\;{\tau}_{23}$ secs.
We may convert this number density of neutrinos into a mass density
${\rho}_{\nu}(0)$ by using the mass derived for a decaying neutrino in the
theory, namely
$27.4\pm 0.2$ eV (Sciama 1993).
We then find that

$${\rho}_{\nu}(0)=2.04{\rm x}10^{-24}\;{\tau}_{23}\;{\rm gm.}{\rm cm}^{-3}$$
$$=0.03\;{\tau}_{23}\;{\rm M}_{\odot}\; {\rm pc}^{-3}\;.$$
We now ask what is the largest observationally permitted value for
${\rho}_{\nu}(0)$, since
we shall soon see that the extragalactic background at $1500 {\AA}$ leads
to a strong lower bound on ${\tau}_{23}$. The mass models of Dehnen and
Binney (1997) provide a
detailed answer to this question (cf. also Gates et al 1995), but a rough
estimate can be derived in the following simple manner. We may obtain
observational constraints
on ${\rho}_{\nu}(0)$ in two ways, by considering (i) estimates for the
total density ${\rho}(0)$ near the sun (the Oort limit) and (ii) the column
densities at
the sun for various matter distributions. The value of the Oort limit is
controversial (eg Kuijken and Gilmore (1991), Bahcall et al 1992). Recent
data from the Hubble
Space Telescope have placed strict limits on the contribution to
${\rho}(0)$ from very faint stars (eg. Gould et al 1996). A reasonable
upper limit on ${\rho}_{\nu}(0)\;,$ derived
from observational estimates of the gravitational force due to ${\rho}(0)
{\rm within}{\sim}300$ pcs of the plane and of the density
${\rho}_{obs}(0)$ of observed material, would be $0.1$ M$_{\odot}\;{\rm
pc}^{-3},$
which would lead to an upper limit on ${\tau}_{23}\; {\rm of}{\sim}3$. In
this connexion we note that Binney et al suggested already in 1987 that all
the dark matter near the sun might
be due to a flattened halo.

A more stringent upper limit on ${\rho}_{\nu}(0)\;,$ and therefore on
${\tau}_{23},$ follows from considering various column densities at the
sun. According to Kuijken and Gilmore (1991) the column density
${\sum}_{1.1}$ of the observed and dark matter combined out to $1.1$ kpc is
$71\pm 6$ M$_{\odot}\; {\rm pc}^{-2}$. Some authors have argued that their
error estimates
should be increased somewhat (eg. Bahcall et al 1994). We therefore follow
Gates et al (1995) and assume that ${\sum}_{1.1} \leq 100$ M$_{\odot}\;{\rm
pc}^{-2}$. For the
total column density ${\sum}_{obs}$ of the observed material at the sun we
adopt ${\sum}_{obs}=40 $M$_{\odot}\;{\rm pc}^{-2}$ (Gould et al 1996).
Hence ${\sum}_{{\nu},1.1}\leq \;60 $M$_{\odot}\;{\rm pc}^{-2}$.
So long as the scale height of the neutrino distribution is much greater
than $1.1$ kpc, we have that $0.03\; {\tau}_{23}$ x$2.2$x$10^3 \leq\; 60$
and so ${\tau}_{23}\leq \;0.9$. In view of the
uncertainties in the values we have adopted we shall suppose that
${\tau}_{23}\leq\;1$. Thus the upper limit on ${\rho}_{\nu}(0)$ is $0.03$
M$_{\odot}\;{\rm pc}^{-3}$.
When we come to consider the background at $1500 \AA$ we shall find that
$\tau_{23}\geq1$, so that the only consistent possibility is
$\tau_{23}\sim1$ and ${\rho}_{\nu}(0)\sim 0.03$ M$_\odot {\rm pc}^{-3}$.
We therefore examine the implications of this value of ${\rho}_{\nu}(0)$
for the flattening of the neutrino halo.

Since a large ${\rho}_{\nu}(0)$ implies a flattened neutrino halo, we begin
by considering the total column density ${\sum}_{rot}$ of a flattened
system required to account for the rotation velocity $v_c$ of the Galaxy
at the sun's position $(R=R_{\odot})$. Binney and Tremaine (1987) give for
this quantity
\begin{eqnarray}
\sum_{rot}&=&\frac{v^2_c}{2{\pi}GR_{\odot}}
\nonumber \\
&=&210 \;{\rm M}_{\odot}\;{\rm pc}^{-2}
\nonumber
\end{eqnarray}
for $v_c=220\; {\rm km}\;{\rm sec}^{-1}$ and $R_{\odot}=8.5 \;{\rm kpc}$.
Hence
$${\sum}_{\nu}=170 \;{\rm M}_{\odot}\;{\rm pc}^{-2}\;,$$
and so the scale height $l_{\nu}$ of the neutrino distribution is given by
$l_{\nu}=2.8$ kpc, which is indeed substantially greater than $1.1$ kpc, so
that our previous discussion is
valid. To derive the implied flattening of the neutrino halo we assume that
the neutrino distribution in the plane has the form
${\rho}_{\nu}(r)={\rho}_{\nu}(0){a^2}/{(a^2+r^2)}$,
with $a=8 \;{\rm kpc}$ (Sciama 1993).Then the "scale-height" in the plane
is ${\pi a}/2$ or $4\pi \;{\rm kpc}$, and so the flattening q is given by
$q=0.2$,
corresponding to the shape $E8$. This simple reasoning is confirmed by the
Dehnen-Binney models which Walter Dehnen kindly extended for me to include
$q=0.3$ and $q=0.2$, for which $\rho_\nu(0)=0.03 {\rm M}_\odot\, 
{\rm pc}^{-3}$ lies just on the edge of the allowed range of
models (cf.\  also Gates et al.\ 1995).

\section{$\tau_{23}$ and the $H\alpha$ Data}
We now show that Reynolds' (1984) $H\alpha$ data, as interpreted by the
decaying neutrino theory, also lead to the result $\tau_{23}\sim 1$. We
consider separately the ionisations produced by decay photons in the opaque
layer of free electrons lying above
 and below the sun out to a distance {\it l} and those produced by decaying
neutrinos lying in the transparent regions outside this layer. A detailed
model of the opaque layer would be complicated, because we should consider
the contribution of both clouds and
the intercloud medium to the opacity. For simplicity we shall assume that
the opaque region corresponds to the electron layer of scale height {\it l}
as derived from the pulsar dispersion data. Reynolds (1991),  Nordgren et
al (1992) and Taylor and Cordes (1993) found that $l \sim 1$ kpc.
Inside this opaque layer every decay photon produces an ionisation in its
vicinity, so along a line of sight normal to the galactic plane there will
be $n_l$ ionisations in this layer, where
$n_l=2.6$x$10^{-13}$x$(0.04)^2$x$6$x$10^{21}$ or $2.5$x$10^6$.
The column density of neutrinos outside the opaque layer corresponds to
$115.5 \;{\rm M}_{\odot}\;{\rm pc}^{-2}$ and so is $5$x$10^{29}$ neutrinos
cm$^{-2}$. The number of ionisations which they produce inside the layer is
reduced by the usual factor $4$ which arises
from an integration over solid angle related to the slab geometry. Hence
they produce ${5{\rm x}10^6}/{(4 \;{\tau}_{23})}$ ionisations
cm$^{-2}$sec$^{-1}$.
Thus
$$\frac {1.25{\rm x}10^6}{\tau_{23}}+2.5{\rm x}10^6=4{\rm x}10^6\;,$$
or
$${\tau}_{23}=0.8\;.$$
Given the simplicity of our model for the opaque layer we round this result
off to ${\tau}_{23}\sim 1,$ which is just compatible with our previous
result $\tau_{23}\leq 1.$

\section{$\tau_{23}$ and the Extragalactic Background at $1500 \AA$}
Some of the earliest lower limits on $\tau_{23}$ were based on
observational estimates of the cosmic background in the far u-v, due to
red-shifted decay photons produced by the cosmological distribution of
neutrinos (Stecker 1980, Kimble et al 1981). As
mentioned in the introduction we here adopt an observed flux of $300\pm 80
$CU at $1500 \AA$. The most recent estimate (Armand et al 1994) for the
contribution due to galaxies at $2000\;{\AA}$ is $40-130$ CU.
The red-shifted contribution from decay photons has recently been
recalculated by Sciama (1991), Overduin et al (1993) and Dodelson and Jubas
(1994). The main uncertainty arises from absorption by dust in the Galaxy.
Allowing a factor $2$ for this absorption, one obtains
$400\;{\tau}_{23}^{-1}$ CU. Given the uncertainties, a reasonable
conclusion is that
$${\tau}_{23}\geq 1.$$
In conjunction with our previous discussion we arrive at a solution which
is just consistent with all the observational constraints, with
$${\tau}_{23}\sim 1\;,$$
$${\rho}_{\nu}(0)\sim 0.03\;{\rm M}_\odot \;{\rm pc}^{-3},$$
$$q\sim 0.2 \;.$$
This solution corresponds to the most flattened possible halo for our
Galaxy (E8).

\section{Conclusions}
We ask in conclusion whether such a large flattening is otherwise
reasonable. I believe that it is. It is noteworthy that another galaxy,
NGC4650A, is observed to have a highly flattened halo. This was deduced
from observations of an outer ring of gas, dust and stars which are on
orbits that are nearly perpendicular to the plane of the flattened central
galaxy, which rotates about its own apparent minor axis (a polar ring
galaxy) These orbits delineate the gravitational potential of the galaxy
outside its central plane. Sackett et al (1994) deduced that the halo of
this galaxy is flattened towards the plane of its central body. They state
that whenever the data were ambiguous they attempted to err on the side of
favouring rounder halos. Still they found for this galaxy that q lies
between $0.3$ and $0.4 (E6-E7)$. I therefore regard the requirement from
the decaying neutrino theory, that the dark halo of our Galaxy is as flat
as it could possibly be, is a reasonable one.

\section{Acknowledgements}
I am very grateful to Walter Dehnen for extending the Dehnen-Binney mass
models of our Galaxy into the extreme regime required by the decaying
neutrino theory, and for helpful discussions of the limits imposed by these
mass models. I am also grateful to Geza Gyuk for his advice.
This work was financially supported by the Ministero dell'Universita' e
della Ricerca Scientifica.

\end{document}